\documentclass[preprint]{revtex4}

\usepackage{graphicx}
\usepackage{dcolumn}
\usepackage{amsmath}

\begin{document}

\title{Riccati equations for bounded radiating systems}

\author{S. D. Maharaj}\email[]{maharaj@ukzn.ac.za}
\author{A. K. Tiwari}
\author{R. Mohanlal}
\author{R. Narain}

\affiliation{Astrophysics and Cosmology Research Unit, School of Mathematics, Statistics and Computer Science, 
University of KwaZulu--Natal, Private Bag X54001, Durban 4000, South Africa}

\begin{abstract}
We systematically analyze the nonlinear partial differential equation that determines the behaviour  of a bounded radiating spherical mass in general relativity. Four categories of solution are possible. These are identified in terms of restrictions on the gravitational potentials. One category of solution can be related to the horizon function transformation which was recently introduced. A Lie symmetry analysis of the resulting Riccati equation shows that several new classes of exact solutions are possible. The relationship between the horizon function, Euclidean star models and other earlier investigations is clarified. \\ \\
\emph{Keywords}: Nonlinear equations; Lie symmetries; radiating matter
\end{abstract}

\maketitle


\section{Introduction}

Radiating stars with strong gravitational fields are an important subject of research because of many applications in astrophysical phenomena. Closed form solutions of the Einstein field equations in the interior, the consistency of the junction conditions at the boundary and matching to the Vaidya radiating exterior are necessary for describing a radiating object in general relativity. Such models provide a valuable insight into the physics of relativistic phenomena in the life history of a relativistic star. Of particular importance are astrophysical processes associated with gravitational collapse and dissipation. These are highlighted in the recent investigations of Sharma and Tikekar \cite{sharma}, Sarwe and Tikekar \cite{sarwe}, Reddy et al \cite{reddy} and Tewari \cite{tewari1,tewari2}. An investigation of the mathematical structure of the relevant differential equations allows for a deeper physical study. Exact solutions of the field equations and the boundary condition are required for modeling the radiating star.  

Several researchers have studied this topic in different physical settings. Most of the earlier treatments placed restrictions on the trajectories of the fluid particles, affecting the acceleration and expansion, and required that the star be shear-free. However it is possible to solve the field equations and the boundary condition at the surface and generate closed form solutions. Exact models with acceleration, expansion and shear were presented by Thirukkanesh et al \cite{thiru3}, Thirukkanesh and Govender \cite{thir1}, Herrera and Santos \cite{hh7}, Govender et al. \cite{hh8} and Govinder and Govender \cite{hh9}. Abebe et al \cite{b12} obtained a generalized class of Euclidean stars, with a barotropic equation of state, using the method of Lie symmetries on differential equations. The usual Lie method may be extended to include exponential Lie symmetry generators which produce the most general exact solutions as recently demonstrated by Mohanlal et al. \cite{Mohan} Another interesting approach was followed by Ivanov \cite{ivo1} in which he introduced a new variable called the horizon function. This method has the advantage of substantially reducing the complexity of the boundary condition and the horizon function can be related to physical features of the radiating star.

In this paper, we discuss the symmetry invariance of the junction condition for a radiating star which is accelerating, expanding and shearing. It has already been proved that the method of Lie theory of extended transforms applied to differential equations is very useful for obtaining exact solutions for a general relativistic star. For example, in the previous study, general radiating stars have been studied by Abebe et al. \cite{b12} and a set of new exact solutions have been found. Recently, Ivanov \cite{ivo1} studied the same problem by introducing a horizon function. This horizon function transforms the junction condition to a simpler form. This motivates us to investigate the Lie symmetries of the transformed equation. Using the invariance condition the transformed equation can be reduced to an ordinary differential equation of the Riccati type. We then investigate the transformed ordinary differential equation to obtain new exact solutions. We also identify the cases when the horizon function transformation does not hold; these additional cases are analyzed.

We investigate the junction conditions for a general relativistic star that is accelerating, expanding and shearing. The governing dynamical equation is a highly nonlinear differential equation. The structure of this investigation is as follows. In Sect. \ref{sect2} we derive the partial differential equation governing the stellar boundary.  To simplify the boundary condition, we  apply the transformation introduced by Ivanov \cite{ivo1} to obtain a simpler partial differential equation. In Sect. \ref{analysis} we analyze the horizon function transformation and consider the conditions for which it is valid. Four categories of solutions are possible. A Lie symmetry analysis of the transformed boundary condition is performed in Sect. \ref{Lie Symmetries}. The symmetry generators of the transformed boundary condition are presented and the reduction variables are constructed. Using the reduction variables we then reduce the transformed equation to a Riccati equation. Exact solutions to the Riccati equation are found in Sect. \ref{newsolutions}. The functional forms of the potentials and horizon functions are derived and presented. We show in Sect. \ref{pastresults} that we can regain known results from our analysis. 
 In Sect. \ref{discussion} we give a brief discussion on the solutions obtained and consider possible physical applications.

%=================================================================================================================================================
\section{The model}\label{sect2}

We are considering the general case of a radiating star with acceleration, expansion and shear. The interior metric has the form
\begin{eqnarray}\label{mod1}
	ds^2=-A^2\,dt^2+B^2\,dr^2+Y^2\left(d\theta^2+\sin^2 \theta\,d\phi^2\right),
\end{eqnarray} 
where $A(r,t),\,B(r,t)$ and $Y(r,t)$ are gravitational functions. In the stellar interior the fluid velocity $u^a$ is comoving and timelike so that $u^a\,u_a=-1$. The heat flux vector $q^a$ is orthogonal to $u^a$ with $q_a\,u^a=0$. An observer moving with velocity $u^a$ measures the energy density $\mu$ and the isotropic pressure $p$.

These quantities have the particular form 
\begin{subequations}\label{mod2}
	\begin{eqnarray}
		u^a&=&\left(\frac{1}{A},0,0,0\right),\\
		q^a&=&\left(0,B\,q,0,0\right),\\
		p&=&\frac{1}{3}\,\left(p_{\parallel}+2p_{\perp}\right),
	\end{eqnarray}
\end{subequations}
where $p_{\parallel}$ is the radial pressure and $p_{\perp}$ is the tangential pressure. We follow the notation and conventions of Abebe et al \cite{b12}. The Einstein field equations in the stellar interior are 
\begin{subequations}\label{mod3}
	\begin{eqnarray}
		\mu&=&\frac{2}{A^2}\frac{\dot{B}}{B}\frac{\dot{Y}}{Y}+\frac{1}{Y^2} +\frac{1}{A^2}\frac{\dot{Y}^2}{Y^2}-\frac{1}{B^2}\left(2 \frac{Y''}{Y}+\frac{Y'^2}{Y^2}-2\frac{B'}{B}\frac{Y'}{Y}\right), \label{mod3.1}  \\
		p_{\parallel}&=&\frac{1}{A^2}\left( -2\frac{\ddot{Y}}{Y}-\frac{\dot{Y}^2}{Y^2}+2\frac{\dot{A}}{A}\frac{\dot{Y}}{Y}\right)+\frac{1}{B^2}\left( \frac{Y'^2}{Y^2}+2\frac{A'}{A}\frac{Y'}{Y}\right)-\frac{1}{Y^2},\label{mod3.2} \\
		p_ {\perp}&=&-\frac{1}{A^2}\left( \frac{\ddot{B}}{B}-\frac{\dot{A}}{A}\frac{\dot{B}}{B}+\frac{\dot{B}}{B}\frac{\dot{Y}}{Y}-\frac{\dot{A}}{A}\frac{\dot{Y}}{Y}+\frac{\ddot{Y}}{Y}\right)\nonumber\\
		&&+\frac{1}{B^2}\left( \frac{A''}{A}-\frac{A'}{A}\frac{B'}{B}+\frac{A'}{A}\frac{Y'}{Y}-\frac{B'}{B}\frac{Y'}{Y}+\frac{Y''}{Y}\right),  \label{mod3.3}\\
		q&=&-\frac{2}{AB}\left(-\frac{\dot{Y}'}{Y} +\frac{\dot{B}}{B}\frac{Y'}{Y}+\frac{A'}{A}\frac{\dot{Y}}{Y}\right),\label{mod3.4} 
	\end{eqnarray}
\end{subequations}
for the spherically symmetric metric \eqref{mod1}.

The exterior of the star is governed by the metric 
\begin{eqnarray}\label{mod4}
	ds^2=-\left(1-\frac{2m\left(v\right)}{R}\right)\,dv^2-2\,d\,v\,dR+R^2\left(d\theta^2+\sin^2 \theta\,d\phi^2\right),
\end{eqnarray}
which is the Vaidya radiating solution. The matching of the metrics \eqref{mod1} and \eqref{mod4} together with continuity of extrinsic curvature lead to a set of junction conditions. These conditions result in the differential equation 
\begin{eqnarray}\label{mod5}
	(p_ \parallel)_{\Sigma}&=&(q)_{\Sigma},
\end{eqnarray} 
at the comoving hypersurface $\Sigma$ which is the boundary of the star. Equations \eqref{mod3.2}, \eqref{mod3.4} and \eqref{mod5} lead to the condition
\begin{eqnarray} \label{mod6} 
	&&2AB^2Y\ddot{Y}+AB^2\dot{Y}^2-2B^2Y\dot{A}\dot{Y}-2ABYA'\dot{Y}+2A^2BY\dot{Y}'-2A^2YA'Y'\nonumber\\
	&&-2A^2Y\dot{B}Y'-A^3Y'^2+A^3B^2=0,
\end{eqnarray}
which is a nonlinear differential equation in the potentials $A,\,B$ and $Y$. Equation (\ref{mod6}) is the main object of study in the model of a radiating stellar object.

Ivanov \cite{ivo1} introduced an interesting transformation that greatly reduces the complexity of (\ref{mod6}). Let us consider
\begin{eqnarray}\label{mod7}
	H=\frac{Y'}{B}+\frac{\dot Y}{A},
\end{eqnarray}
which is the so called horizon function. With the help of \eqref{mod7} the boundary condition \eqref{mod6} can be written as
\begin{eqnarray} \label{mod8}
	\dot{H}=\left(\frac{A}{2Y}+\frac{A'}{Y'}\right)\,H^2-\left(\frac{A}{Y}+\frac{A'}{Y'}\right)\,\frac{\dot Y}{A}\,H-\frac{A}{2Y},
\end{eqnarray}
subject to the restrictions $\dot{Y}\neq A$ and $ Y^{\prime}\neq B$.

This is an equation with variables $A$, $H$ and $Y$; the potential $B$ has been replaced by the new quantity $H$. It is clear that (\ref{mod8}) is a simpler equation than (\ref{mod6}). Equation (\ref{mod8}) has been integrated for geodesic motion by Ivanov \cite{ivo2}, and in particular cases with nongeodesic motion and shear in \cite{ivo1}.

 %=============================================================================================================================

\section{Analysis}\label{analysis}
We now analyze the transformation \eqref{mod7}. It is important to observe that equation \eqref{mod8} is not equivalent to the original equation \eqref{mod6}. The transformation \eqref{mod7}, and consequently \eqref{mod8}, is not valid for $Y^{\prime}= B$ or $\dot{Y}= A$. Therefore the analysis will consist of four categories namely $Y^{\prime}=B$, $\dot{Y}=A$, $Y^{\prime}=B$ with $\dot{Y}=A$, and $Y^{\prime}\neq B$ with $\dot{Y}\neq A$.

\subsection{Category $1$: $	B=Y^{\prime}$}\label{analysis1}
When $B=Y^{\prime}$ the transformation \eqref{mod7} does not hold and we cannot use \eqref{mod8}. In this case we have to analyze the original equation \eqref{mod6}. Using $B=Y^{\prime}$ then equation \eqref{mod6} can be written as
\begin{equation}\label{euclideanstar}
\frac{\ddot Y}{Y} + \frac{1}{2}\left(\frac{\dot Y}{Y}\right)^2 - \frac{\dot A}{A}\frac{\dot Y}{Y} - \frac{(A + \dot{Y})A'}{YY'} = 0
\end{equation}
Equation \eqref{euclideanstar} corresponds to an Euclidean star. Some exact solutions to \eqref{euclideanstar} were found by Govender et al. \cite{hh8} and Govinder and Govender \cite{hh9}.

\subsection{Category $2$: $A=\dot{Y}$}\label{analysis2}
When $A=\dot{Y}$ then \eqref{mod7} is not valid and we cannot use \eqref{mod8}. As in Sect. \ref{analysis1}. we have to consider the original equation \eqref{mod6} which becomes
\begin{equation}\label{category2}
\dot{B}-\frac{\dot{Y}}{YY^{\prime}}B^2+\dot{Y}^{\prime}+\frac{\dot{Y}Y^{\prime}}{2Y}=0,
\end{equation}
where we have written \eqref{category2} as a Riccati equation in $B$.

It should be observed that the condition $A=\dot{Y}$ is a special case of a relationship derived by Thirrukanesh et al. \cite{thiru3} in their investigation of a radiating star. To solve \eqref{category2} we follow their approach  and assume $Y$ is a separable function of the form
\begin{equation}\label{ansatz1}
Y=K(r)C(t),
\end{equation}
where $K(r)$ and $C(t)$ are arbitrary functions of their respective arguments. Then \eqref{category2} becomes
\begin{equation}\label{beqn}
\dot{B}-\frac{\dot{C}(t)}{K^{\prime}(r)C(t)^{2}} B^{2}+\frac{3K^{\prime}\dot{C}(t)}{2}=0.
\end{equation}
Using the transformation
\begin{equation}\label{tra1}
V(r,t)=\exp\left(-\frac{1}{K^{\prime}(r)}\int\frac{\dot{C}(t)}{C(t)^2}B(r,t)dt\right),
\end{equation}
we can write \eqref{beqn} as the homogeneous second order ordinary differential equation 
\begin{equation}\label{so1}
\ddot{V}-\left(\frac{\ddot{C}}{\dot{C}}-\frac{2\dot{C}}{C}\right)\dot{V}-\frac{3\dot{C}^2}{2C^2}V=0.
\end{equation}
It should be noted that \eqref{so1} can be transformed into an  equation with constant coefficients by making use of the change of independent variable 
\begin{equation}\label{tra2}
y=\log(C(t)).
\end{equation}
The resulting equation can be written as
\begin{equation}\label{so2}
V_{yy}+V_{y}-\frac{3}{2}V=0.
\end{equation}
Solving equation \eqref{so2} we obtain
\begin{equation}\label{int1}
V=C(t)^{-\left(\frac{1+\sqrt{7}}{2}\right)}(D(r)+U(r)C(t)^{\sqrt{7}}),
\end{equation}
where $D(r)$ and $U(r)$ are arbitrary functions of integration. Setting $D(r)=1$ and 
 inverting the transformations \eqref{tra2} and \eqref{tra1}, we find that
\begin{equation}
B= K'(r)C(t)\left(\frac{1-\sqrt{7}}{2}+\frac{\sqrt{7}}{1+U(r)C(t)^{\sqrt{7}}}\right).
\end{equation} The remaining potential is simply
\begin{equation}
	A=K(r)\dot{C}(t).
\end{equation}
%========================================================================================================================
\subsection{Category $3$: $Y^{\prime}=B,\;\dot{Y}=A$}
 For  $Y^{\prime}=B$ and $\dot{Y}=A$ the transformation \eqref{mod7} is not valid and \eqref{mod8} cannot be used. The original boundary condition \eqref{cu6} has to be analyzed. Under the conditions $Y^{\prime}=B$ and $\dot{Y}=A$ equation \eqref{cu6} simplifies to 
 \begin{equation}
 \dot{Y}^{\prime}-\frac{\dot{Y}Y^{\prime}}{4Y}=0.
 \end{equation}
We can integrate this to obtain
\begin{equation}\label{newpot1}
Y=(C(t)+D(r))^{\frac{4}{3}},
\end{equation}
where $C(t)$ and $D(r)$ are arbitrary functions of their respective arguments. The remaining potentials become
\begin{subequations}\label{newpot2}
	\begin{eqnarray}
	A&=&\frac{4\dot{C}(t)}{3}(C(t)+D(r))^{\frac{1}{3}},\\
	B&=&\frac{4D^{\prime}(r)}{3}(C(t)+D(r))^{\frac{1}{3}}.
	\end{eqnarray}
\end{subequations}
The potentials \eqref{newpot1} and \eqref{newpot2} have simple forms. The matter $\mu$, $p_{\parallel}$ and $q$ are nonzero.  The tangential pressure $p_{\perp}=0$  in this case; the heat flux $q$ is negative
which may be interpreted as inflow of energy from the surrounding medium across the stellar boundary.

\subsection{Category $4$: $Y^{\prime}\neq B$, $\dot{Y}\neq A$}\label{analysis3}
When $Y^{\prime}\neq B$ and $\dot{Y}\neq A$ then the transformation \eqref{mod7} holds and we need to solve the simpler equation \eqref{mod8}. This is done in the next section.
 
 %=========================================================================================================================
 
  \section{Lie Symmetries}\label{Lie Symmetries}
  
  The transformed equation \eqref{mod8} is a complicated partial differential equation in three dependent variables and cannot be solved easily. We therefore make use of Lie symmetries to find solutions to \eqref{mod8}.
 The theory of Lie symmetries \cite{ibra,olver,step} is a powerful tool for finding exact solutions to differential equations. It is based on the point transformations that leave a differential equation  form invariant. The main idea in finding the Lie group of symmetry transformations is to determine the infinitesimal generators of the groups. This can be achieved by deriving the determining characteristic equations; when these equations are solved bases of the generating vector fields are are obtained. Group theoretic methods have been used to produce exact solutions of the Einstein field equations in four and higher dimensions. Using the Lie symmetry method  Govinder and Govender \cite{hh9} produced a general class of Euclidean star models. In keeping with this approach Abebe et al. \cite{b10} generated exact solutions describing conformally flat and radiating stars. The technique of Lie symmetries is algorithmic and several  software packages such as Symmetry, Sym, PROGRAM LIE and Reduce have been developed. In this investigation, we consider the software package PROGRAM LIE  which is described by Head \cite{head}, for finding the Lie symmetries of \eqref{mod8}.
 
  In \cite{b12} it is shown that \eqref{mod6} admits the symmetry generators
  \begin{subequations}\label{abebesym}
  	\begin{eqnarray}
  	\tilde{X}_{1}&=&A\dot{\beta}(t)\frac{\partial}{\partial A}-\beta(t)\frac{\partial}{\partial t},\\
  	\tilde{X}_{2}&=&B\alpha^{\prime}(r)\frac{\partial}{\partial B}-\alpha(r)\frac{\partial}{\partial r},\\
  	\tilde{X}_{1}&=&A\frac{\partial}{\partial A}+B\frac{\partial}{\partial B}+Y\frac{\partial}{\partial Y}.
  	\end{eqnarray}
  \end{subequations}
  New exact solutions to \eqref{mod6} where found by Abebe et al. \cite{b12} via the symmetries \eqref{abebesym}.
  
  With the help of PROGRAM Lie we find that equation \eqref{mod8} admits the symmetry vectors
  \begin{subequations}\label{cu1}
  	\begin{eqnarray}
  	X_1&=&F_{1}(t)\frac{\partial}{\partial t}-A\dot{F_{1}}(t)\,\frac{\partial}{\partial A},\\
  	X_2&=&F_{2}(r)\frac{\partial}{\partial r},\\
  	X_3&=&Y\frac{\partial}{\partial Y}+A\,\frac{\partial}{\partial A},
  	\end{eqnarray}
  \end{subequations}
  where $F_{1}(t)$ and $F_{2}(r)$ are functions of their respective arguments and dots denote differentiation with respect to $t$.   The symmetry generators \eqref{cu1} satisfy the commutation relations
  \begin{subequations}\label{commrel}
  	\begin{eqnarray}
  	&&[X_{1},X_{2}]=0\\
  	&&[X_{1},X_{3}]=0\\
  	&&[X_{2},X_{3}]=0.
  	\end{eqnarray}
  \end{subequations}
  From \eqref{commrel} we deduce that the symmetry vectors \eqref{cu1} generate an Abelian Lie algebra. Consequently, each linear combination of \eqref{cu1} will generate its own class of solutions that cannot be transformed into solutions obtained via any other linear combination.
  
   It should be noted that this symmetry analysis is fundamentally different from the analysis of Abebe et al. \cite{b12} for the boundary condition \eqref{mod6}. The transformation \eqref{mod7} removes the variable $B$ and results in a simpler equation, and as mentioned in Sect. \ref{analysis} the transformation is not general.

For the analysis of \eqref{mod8} we consider the most general combination
 \begin{equation}\label{gencom}
 X=a_{1}X_{1}+a_{2}X_{2}+a_{3}X_{3},
 \end{equation}
 where $a_{1},a_{2}\neq0$ and $a_{3}$ are real constants.  In order to reduce the simpler equation \eqref{mod8} we need to construct the reduction variables by solving the associated Lagrange's system of \eqref{gencom} which is
 \begin{eqnarray}\label{cu3}
 \dfrac{dt}{a_1 F_{1}(t)}=\dfrac{dr}{a_2F_{2}(r)}=\dfrac{dA}{A(a_3-a_1\dot{F_{1}}(t))}=\dfrac{dH}{0}=\dfrac{dY}{a_3Y}.
 \end{eqnarray}
 Integrating \eqref{cu3} yields the reduction variables
 \begin{subequations}\label{cu4}
 	\begin{eqnarray}
 	z&=&\frac{1}{a_1}\int\frac{dt}{F_{1}(t)}-\frac{1}{a_2}\int\frac{dr}{F_{2}(r)},\\
 	A&=&\frac{f(z) e^{\frac{a_3}{a_1}\int\frac{dt}{F_{1}(t)}}}{F_{1}(t)},\\
 	H&=&g(z),\\
 	Y&=&h(z)e^{\frac{a_3}{a_1}\int\frac{dt}{F_{1}(t)}},
 	\end{eqnarray}
 	where $z$ is the new independent variable and $f(z)$, $g(z)$ and $h(z)$ are arbitrary functions which constitute the new dependent variables.  
 \end{subequations}
 
  With the help of equations \eqref{mod7} and \eqref{cu4} the  potential $B$ can be written as 
  \begin{eqnarray}\label{cu4.1}
  B=-\frac{a_1f(z)h'(z)e^{\frac{a_3}{a_1}\int\frac{dt}{F_{1}(t)}}}{a_2F_{2}(r) \left(a_1\,f(z)\,g(z)-a_3\,h(z)-h'(z)\right)}.
  \end{eqnarray}
  We have shown that the potential $B$ is not a free function but depends on the function $h(z)$ which is present in the potential $Y$. This is a consequence of the transformation \eqref{mod7} and the existence of a Lie symmetry. This restriction does not arise in the solutions in Sect. \ref{analysis1}, Sect. \ref{analysis2} or the models of Abebe et al. \cite{b12}. Therefore the transformation \eqref{mod7} may lead to new solutions, not easily obtainable in its absence, but other solutions may also exist.
   
  Substituting \eqref{cu4} into equation \eqref{mod8} we obtain, after simplification, the reduced equation
  \begin{eqnarray}\label{cu5}
  	g'(z)+P(z)g(z)^2+Q(z)g(z)+R(z)=0,
  \end{eqnarray} 
  where
  \begin{subequations}\label{cu6}
  	\begin{eqnarray}
  		P&=&-\frac{a_1 }{2}\left(\frac{2 f'(z)}{h'(z)}+\frac{f(z)}{h(z)}\right),\\
  		Q&=&\frac{\left(h'(z)+a_3h(z)\right) \left(h(z) f'(z)+f(z) h'(z)\right)}{f(z) h(z) h'(z)},\\
  		R&=&\frac{a_1 f(z)}{2 h(z)},
  	\end{eqnarray}
  \end{subequations}
  primes now denote differentiation with respect to the variable $z$.
  We have chosen to write the reduced equation \eqref{cu5} as an equation in $g(z)$. Equation \eqref{cu5} is a Riccati equation in $g(z)$ and cannot be solved in general. However, for certain restrictions on the coefficient functions $P(z)$, $Q(z)$ and $R(z)$, \eqref{cu5} simplifies and possibly can be solved. This is done in the next section.

%========================================================================================================================================
\section{New solutions}\label{newsolutions}
In this section we place particular restrictions on each of the coefficient functions in \eqref{cu6} to seek solutions to \eqref{cu5}. We consider each restriction in turn.

\subsection{Case $1$: $P(z)=0$}\label{case1}
In this case \eqref{cu5} reduces to a linear equation. The condition $P(z)=0$ implies that $f(z)$ and $h(z)$  must satisfy the differential equation
\begin{equation}\label{nr1}
\frac{h'}{h}+2\frac{f'}{f}=0.
\end{equation}
We can integrate \eqref{nr1} and find
\begin{equation}\label{nr2}
h(z)=\frac{c_1}{f(z)^2},
\end{equation}
where $c_{1}$ is a constant of integration. Substituting \eqref{nr2} into \eqref{cu5} we obtain 
\begin{equation}\label{nr1.1}
g'(z)+\left(\frac{a_3}{2}-\frac{f'(z)}{f(z)}\right)g(z)+\frac{a_1f(z)^3}{2 c_1}=0.
\end{equation}
We can solve equation \eqref{nr1.1} to obtain the general solution
\begin{equation}\label{gsol}
g(z)=f(z)e^{-\frac{1}{2}a_3 z}\left(c_2-\frac{a_1}{2c_1}\int_1^z  f(w)^2 e^{\frac{1}{2} a_3 w}dw\right),
\end{equation}
where $c_{2}$ is a constant of integration.

With the help of equations \eqref{cu4}, \eqref{cu4.1}, \eqref{nr2} and \eqref{gsol} the horizon function becomes
\begin{equation}
H=f(z)e^{-\frac{1}{2}a_3 z}\left(c_2-\frac{a_1}{2c_1}\int_1^z  f(w)^2 e^{\frac{1}{2} a_3 w}dw\right),
\end{equation}
and the potentials are
\begin{subequations}
	\begin{eqnarray}
	A&=&\frac{f(z)e^{\frac{a_3}{a_1}\int \frac{dt}{F_{1}(t)}}}{F_1(t)},\\
	B&=&\frac{2 a_1 c_1 f(z) f'(z)e^{\frac{a_3}{a_1}\int \frac{ dt}{F_{1}(t)}}}{a_2 F_{2}(r) \left(2 c_1 f'(z)-a_3 c_1 f(z)+a_1 e^{-\frac{1}{2}a_3 z} f(z)^5 \left(c_2-\frac{a_1}{2 c_1} \int e^{\frac{a_3 w}{2}} f(w)^2dw\right)\right)},\\
	Y&=&\frac{c_1e^{{\frac{a_3}{a_1}\int\frac{ dt}{F_1(t)}}}}{f(z)^2}.
	\end{eqnarray}
\end{subequations}

%====================================================================================================================

\subsection{Case $2$: $Q(z)=0$ }
In this case \eqref{cu5} becomes a simpler Riccati equation. With the restriction $Q(z)=0$, we have the constraints
\begin{equation}\label{subsub1}
\frac{h'}{h}+\frac{f'}{f}=0,
\end{equation}
or 
\begin{equation}\label{subsub2}
h'+a_{3}h=0.
\end{equation}
We consider each of these conditions separately.

\subsubsection{$\frac{h'}{h}+\frac{f'}{f}=0$}
Equation \eqref{subsub1} can be solved to yield
\begin{equation}\label{nr7.1}
	h(z)=\frac{c_1}{f(z)},
\end{equation}
where $c_{1}$ is a constant of integration. Substituting \eqref{nr7.1} into  equation \eqref{cu5} yields
\begin{equation}\label{nr7.1.1}
g'(z)+\frac{a_1f(z)^2}{2c_1}g(z)^2+\frac{a_1f(z)^2}{2c_1}=0.
\end{equation}
Using the transformation
\begin{equation}\label{tra3}
u(z)=\exp\left(\frac{a_1}{2c_1}\int f(z)^2g(z)dz\right),
\end{equation}
we can write equation \eqref{nr7.1.1} as the second order homogeneous equation
\begin{equation}\label{so3}
u''-\frac{2f'}{f}u'+\frac{a_1^2f^4}{4c_1^2}u=0.
\end{equation}
We can write equation \eqref{so3} as an equation with constant coefficients by making use of the change of independent variable
\begin{equation}\label{tra4}
y=\int f(z)^2dz.
\end{equation}
 This results in the simpler equation
\begin{equation}\label{so4}
u_{yy}+\frac{a_1^2}{4c_1^2}u=0.
\end{equation}
Solving \eqref{so4} we find
\begin{equation}
u=d_{1}\cos\left(\frac{a_1 z}{2c_1}\right)+d_2\sin\left(\frac{a_1 z}{2c_1}\right),
\end{equation}
where $d_{1}$ and $d_{2}$ are integration constants. Setting $d_{1}=1$, $d_{2}=\tan c_2$ and inverting the transformations \eqref{tra4} and \eqref{tra3} we obtain the solution
\begin{equation}\label{gsol7.1.1}
g(z)=\tan\left(c_2-\frac{a_1}{2 c_1}\,\int^{z}_{1} f(w)^2 dw\right),
\end{equation}
where $c_{2}$ is an arbitrary constant. 

Substituting \eqref{nr7.1} and \eqref{gsol7.1.1} into equations \eqref{cu4} and \eqref{cu4.1} we find that 
\begin{equation}
H=e^{\frac{a_3}{a_1}\int \frac{dt}{F_{1}(t)}}\,\tan\left(c_2-\frac{a_1}{2 c_1}\int^{z}_{1} f(w)^2dw\right).
\end{equation} 
The potentials become
\begin{subequations}
	\begin{eqnarray}
		A&=&\frac{f(z)e^{\frac{a_3}{a_1}\int \frac{dt}{F_{1}(t)}}}{F_{1}(t)},\\
		B&=&-\frac{a_1 c_1 f(z) f'(z) e^{ \frac{a_{3}}{a_{1}}\int\frac{dt}{F_1(t)}}}{a_2 F_2(r) \left(a_1 f(z)^3 \tan \left(c_2-\frac{a_1}{2 c_1}\int^{w}_{1} f(w)^2 dw\right)-c_1 f'(z)+a_3 c_1 f(z)\right)},\\
		Y&=&\frac{c_1e^{\frac{a_3}{a_1}\int \frac{dt}{F_{1}(t)}}}{f(z)}.
	\end{eqnarray}
\end{subequations}

\subsubsection{$h'+a_{3}h=0$}
For the condition $h'+a_{3}h=0$ to be satisfied, the function  $h$ has the form
\begin{equation}\label{nr7.2}
h(z)=c_1 e^{-a_3 z},
\end{equation}
where $c_{1}$ is a constant of integration. Then equation \eqref{cu5} can be written as
\begin{equation}\label{friccati}
f'(z)+a_{3}\left(\frac{1-g(z)^2}{2g(z)^2}\right)f(z)+\frac{a_{3}c_{1}e^{-a_{3}z}g'(z)}{a_{1}g(z)^2}=0.
\end{equation}
We have  written \eqref{friccati} as a linear equation in $f(z)$ as this particular form proves to be most useful. Solving \eqref{friccati} for $f(z)$ we find the general solution
\begin{equation}\label{friccatisol}
f(z)=\frac{1}{a_1}e^{\frac{ a_3}{2}\int^{z}_{1}\left(1-\frac{1}{g(w)^2}\right)dw}\left(a_1 c_2-a_3 c_1 \int^{z}_{1} {\frac{g'(w)}{g(w)^2}e^{-a_3 w}  e^{-\frac{ a_3}{2}\int^{w}_{1} \left(1-\frac{1}{g(x)^2}\right)dx}} dw\right),
\end{equation}
where $c_{2}$ is a constant of integration. Note that if we write \eqref{friccati} as a Riccati equation in $g(z)$, then we cannot obtain exact solutions  in $g(z)$ for particular functional choices of $f(z)$.

Using  \eqref{nr7.2} and \eqref{friccatisol} in conjunction with equations \eqref{cu4} and \eqref{cu4.1} the horizon function becomes
\begin{equation}
H=g(z),
\end{equation}
and the potentials are
\begin{subequations}\label{nr17}
	\begin{eqnarray}
	A&=&e^{\frac{1}{2} a_3 \int^{z}_{1} \left(1-\frac{1}{g(w)^2}\right)dw} \left(a_1 c_2-a_3 c_1 \int^{z}_{1}{\frac{g'(w)}{g(w)^2}e^{-a_3 w}  e^{\frac{1}{2} a_3 \int^{w}_{1} \left(1-\frac{1}{g(x)^2}\right)dx}}dw\right)\frac{e^{\frac{a_3}{a_1}\int \frac{dt}{\beta(t)}}}{a_1\beta(t)},\\
	B&=&\frac{a_3 c_1 e^{-a_3 z}e^{\frac{a_3}{a_1}\,\int\frac{dt}{\beta(t)}}}{a_2F_2(r)g(z)},\\
	Y&=&c_1e^{-a_3 z}e^{\frac{a_3}{a_1}\,\int \frac{dt}{\beta(t)}}.
	\end{eqnarray}
\end{subequations}

%=========================================================================================================================

\subsection{Category $3$: $R(z)=k$}
The condition $R(z)=k$, $k\neq 0$ is a real constant, leads to the relation
\begin{equation}\label{nr19}
h(z)=\frac{a_{1}f(z)}{k}.
\end{equation}
Substituting \eqref{nr19} into \eqref{cu5} and writing it as a differential equation in $f(z)$ we obtain a linear  equation
\begin{equation}\label{nr19.1}
f'(z)+\left(a_{3}+\frac{g'(z)}{2g(z)}-\frac{3kg(z)}{2}+\frac{k}{2g(z)}\right)f(z)=0.
\end{equation}
Solving \eqref{nr19.1} we find that
\begin{equation}\label{nr19.1sol}
f(z)=\frac{c_{1}e^{-a_3 z}}{\sqrt{g(z)}}\exp \left(-\frac{1}{2} \int^{z}_{1} \frac{k-3kg(w)^2}{g(w)}dw\right),
\end{equation}
where $c_{1}$ is a constant of integration. Using \eqref{nr19.1sol} in \eqref{nr19} yields
\begin{equation}\label{nr19.1sol1}
h(z)=\frac{a_{1}c_{1}e^{-a_3 z}}{2\,k\,\sqrt{g(z)}}\exp \left(-\frac{1}{2} \int^{z}_{1} \frac{k-3kg(w)^2}{g(w)}dw\right).
\end{equation}

We can use \eqref{nr19.1sol} and \eqref{nr19.1sol1} in conjunction with \eqref{cu4} and \eqref{cu4.1} to obtain the horizon function
\begin{equation}
	H=g(z) e^{\frac{a_3}{a_1}\int \frac{dt}{F_{1}(t)}},
\end{equation}
and we find that the potentials become
\begin{subequations}\label{nr23}
	\begin{eqnarray}
	A&=&\frac{2c_1ke^{-a_3z}e^{ \frac{a_{3}}{a_{1}}\int \frac{dt}{F_{1}(t)}}}{a_1 F_{1}(t)\sqrt{g(z)}}\exp \left(-\frac{1}{2} \int^{z}_{1} \frac{k-3kg(w)^2}{g(w)}dw\right),\\
B&=&\exp \left(-\frac{1}{2} \int^{z}_{1} \frac{k-3kg(w)^2}{g(w)}dw\right)\nonumber\\
&&\times\frac{2 c_1 ke^{-a_3 z} e^{\frac{a_3}{a_1}\,\int \frac{dt}{F_{1}(t)}}\left(g'(z)-3 k g(z)^2+2a_3g(z)+k\right)}{a_2F_{2}(r) \left(g'(z)+k g(z)^2+k\right)\sqrt{g(z)}} ,\\
	Y&=&\frac{c_1e^{-a_3 z}e^{\frac{a_3}{a_1}\,\int \frac{1}{\beta(t)} \, dt}}{\sqrt{g(z)}} \exp \left(-\frac{1}{2} \int^{z}_{1} \frac{k-3kg(w)^2}{g(w)}dw\right).
	\end{eqnarray}
\end{subequations}

%=================================================================================================

\section{Recovering known results}\label{pastresults}
In this section we show that the results of this paper can be related to the earlier results of Abebe et al. \cite{b12}. In the symmetry analysis \cite{b12} of equation \eqref{mod6}  exact solutions were obtained by reducing the boundary condition via the symmetries \eqref{abebesym}.

Consider the transformation
\begin{equation}\label{abr2}
g(z)=\frac{h'(z) (b f(z)+M(z))}{b f(z) M(z)},
\end{equation}
where $M(z)$ is an arbitrary function of $z$. Using \eqref{abr2} in equations \eqref{cu4} and \eqref{cu4.1} we find that the potentials become
\begin{subequations}\label{po1.1}
	\begin{eqnarray}
	A&=&\frac{f(z) e^{\frac{a_3}{a_1}\int\frac{dt}{F_{1}(t)}}}{F_{1}(t)},\\
	B&=&\frac{a_{1}e^{\frac{a_3}{a_1}\int\frac{dt}{F_{1}(t)}}f(z)h'(z)M(z)}{a_{2}F_{2}(r)(a_{3}h(z)M(z)-a_{1}f(z)h'(z))},\\
	Y&=&h(z)e^{\frac{a_3}{a_1}\int\frac{dt}{F_{1}(t)}}.
	\end{eqnarray}
\end{subequations}
If we set $F_{1}(t)=\beta(t)$, $F_{2}(r)=\alpha(r)$ and impose the parametric restrictions $a_{3}=0$, $a_{1}=b$ and $a_{2}=-1$, then
\begin{equation}\label{newz}
z=\frac{1}{b}\int\frac{dt}{\beta(t)}+\int\frac{dr}{\alpha(r)}.
\end{equation}
With the above assumptions the potentials \eqref{po1.1} simplify to
\begin{subequations}\label{po1.2}
	\begin{eqnarray}
	A&=&\frac{f(z)}{\beta(t)},\\
	B&=&\frac{M(z)}{\alpha(r)},\\
	Y&=&h(z).
	\end{eqnarray}
\end{subequations}
  Furthermore, if we assume $f(z)=c M(z)$, where $c$ is a nonzero real constant, and using \eqref{abr2}, then equation \eqref{cu5} can be written as
\begin{eqnarray}\label{arb3}
M'(z)-\left(\frac{b^2 c^2 }{2 (b c+1)^2 h(z) h'(z)}\right)M(z)^3+\left(\frac{(b c-1) h'(z)^2-2 h(z) h''(z)}{2 (b c+1) h(z) h'(z)}\right)M(z)=0.
\end{eqnarray}
This is a Bernoulli equation and has the same form as equation $(22)$ that arises in the generalized Euclidean star model of Abebe et al. \cite{b12}. Integration leads to the potentials
\begin{subequations}\label{po1.3}
	\begin{eqnarray}
	A&=&\frac{ch(z)^{\frac{1-c\,b}{2(1+b\,c)}}h'(z)^{\frac{1}{1+b\,c}}}
	{\beta(t)\left(c_{1}-\frac{c^2b^2}{\left(1+bc\right)^2}\int^{z}_{1}{h(w)^{\frac{-2\,b\,c}{1+b\,c}}h'(w)^{\frac{1-b\,c}{1+b\,c}}dw}\right)^{\frac{1}{2}}},\\
	B&=&\frac{h(z)^{\frac{1-c\,b}{2(1+b\,c)}}h'(z)^{\frac{1}{1+b\,c}}}
	{\alpha(r)\left(c_{1}-\frac{c^2b^2}{\left(1+bc\right)^2}\int^{z}_{1}{h(w)^{\frac{-2\,b\,c}{1+b\,c}}h'(w)^{\frac{1-b\,c}{1+b\,c}}dw}\right)^{\frac{1}{2}}},\\
	Y&=&h(z).
	\end{eqnarray}
\end{subequations}
The potentials \eqref{po1.3} have the same functional form as the generalized Euclidean star. However, note that the results of this investigation is not equivalent to the results in \cite{b12}. In the treatment \cite{b12} Euclidean star models are regained from the generalized Euclidean star by setting $B=Y^{\prime}$. For the class of solutions found in Sect. \ref{newsolutions} we have the restriction $B\neq Y^{\prime}$ for the transformation \eqref{mod7} to be well defined. Therefore, Euclidean star models cannot be regained from the solutions obtained in Sect. \ref{newsolutions}.
%===============================================================================================

%==================================================================================================

\section{Discussion}\label{discussion}

We have systematically studied the fundamental dynamical equation that governs the evolution of a radiating star in general relativity. In particular we have investigated the role of the horizon function transformation introduced by Ivanov \cite{ivo1} which has the advantage of simplifying the nonlinear boundary condition to a simpler Riccati equation. A Lie symmetry analysis of the Riccati equation shows that there are several new families of exact solutions. We have also identified the cases when the horizon function transformation is not applicable; three other categories of solution are possible. These are identified and discussed. Our analysis highlights the role of the horizon function in the integration of the boundary condition and shows its value in identifying   new classes of exact solutions. The relationship between the horizon function and the generalized Euclidean star models of Abebe et al. \cite{b12} has been clarified. The transformation \eqref{mod7} does not allow us to regain the Euclidean star models with $Y^{\prime}=B$; this case has to be treated separately. The fact that we show that the transformation \eqref{mod7} does not allow the gravitational potential $B$ to be a free function indicates that the horizon function has limitations; the potential $B$ is specified by equation \eqref{cu4.1}.

The advantage of the transformation \eqref{mod7} is that it introduces the horizon function $H$ which has several useful physical features. The function $H$ may be associated with the mass $m$, the redshift $Z$, the surface luminosity $\Lambda$, the energy density $\mu$ at the stellar boundary $\Sigma$, and the luminosity at infinity $\Lambda_{\infty}$. These quantities are given by 
\begin{subequations}
	\begin{eqnarray}
	m_{\Sigma}&=&\frac{Y}{2}\left(1-H^{2}+\frac{2\dot{Y}}{A}H\right),\\
	\mu&=&\frac{2m^{\prime}}{Y^{2}Y^{\prime}}-\frac{qB^{2}Y^{2}\dot{Y}}{AY^{\prime}},\\
	Z_{\Sigma}&=&\frac{1}{H}-1,\\
	\Lambda_{\Sigma}&=&\frac{1}{2}qBY^{2},\\
	\Lambda_{\infty}&=&H^{2}\Lambda_{\Sigma}.
	\end{eqnarray}
\end{subequations} 
For the various families of solution found in this paper we have analytic expressions for the potentials and the horizon function $H$. This makes it possible to obtain explicit forms for the physical quantities above and the matter variables. Consequently a physical analysis of the geometrical and physical quantities can be performed to obtain a complete dynamical model of a radiating star in general relativity.

\begin{acknowledgements}
 SDM acknowledges that this work is based upon research supported by the South African Research Chair Initiative of the Department of Science and Technology and the National Research Foundation. AKT, RM and RN thank the  University of KwaZulu-Natal and National Research Foundation  for financial support.
\end{acknowledgements}


\begin{thebibliography}{}
 

	
	\bibitem{sharma}  R. Sharma and R. Tikekar, Gen. Relativ. Gravit. \textbf{44}, 2503 (2012).
	
	\bibitem{sarwe} S. Sarwe and R. Tikekar, Int. J. Mod. Phys. D \textbf{19},  1889 (2010).
	
	\bibitem{reddy} K. P. Reddy,  M. Govender and  S. D. Maharaj, Gen. Relativ. Gravit. \textbf{47}, 35 (2015).
	
	\bibitem{tewari1} B. C. Tewari, Astrophys. Space. Sci. \textbf{342}, 73 (2012).
	
	\bibitem{tewari2} B. C. Tewari, Gen. Relativ. Gravit. \textbf{45}, 1547 (2013).
	
	\bibitem{thiru3}  S. Thirukkanesh, S. S. Rajah and S. D. Maharaj, J. Math Phys. \textbf{53}, 032506 (2012).
	
	\bibitem{thir1}  S. Thirukkanesh and M. Govender, Int. J. Mod. Phys. D \textbf{22}, 1350049 (2013).
	
	\bibitem{hh7} L. Herrera and  N. O. Santos, Gen. Relativ. Gravit. \textbf{42}, 2383 (2010).
	\bibitem{hh8} G. Govender, M. Govender,  K. S. Govinder,  Int. J. Mod. Phys. D  \textbf{19}, 1773 (2010).
	\bibitem{hh9}  K. S. Govinder and M. Govender,  Gen. Relativ. Gravit. \textbf{44}, 147 (2012).
	
	\bibitem{b12}  G. Z. Abebe,   S. D. Maharaj and K. S. Govinder, Gen. Relativ. Gravit.  \textbf{46}, 1733 (2014).
	\bibitem{Mohan} R. Mohanlal, S. D. Maharaj, Ajey. K. Tiwari and R. Narain, Gen. Relativ. Gravit. \textbf{48}, 87 (2016).
	\bibitem{ivo1} B. V. Ivanov, Int. J. Mod. Phys. D \textbf{25}, 1650049 (2016).
	
	\bibitem{ivo2} B. V. Ivanov, Astrophys. Space Sci. \textbf{361}, 18 (2016).
	
	\bibitem{ibra}
	N. H. Ibragimov, Elementary Lie Group Analysis and Ordinary Differential Equations (Wiley,  New York, 1999).
	
	\bibitem{olver}
	P. J. Olver,  Applications of Lie Groups to Differential Equations (Springer-Verlag, New York, 1986).
	
	\bibitem{step}
	H. Stephani,  Differential Equations: Their Solutions Using Symmetries (Cambridge University Press, Cambridge, 1989).
	
	
	
	\bibitem{b10} G. Z. Abebe,  K. S. Govinder and S. D. Maharaj,  Int. J. Theor. Phys. \textbf{52}, 3244 (2013).
	
	\bibitem{head} A. K. Head, Comput. Phys. Commun. \textbf{71}, 241 (1993).

 \end{thebibliography}
\end{document}